\documentclass[runningheads]{llncs}
\usepackage[T1]{fontenc}

\usepackage{graphicx}
\usepackage{balance} 
\usepackage{algorithm}
\usepackage{algorithmic}
\usepackage{amsmath}
\usepackage{bm}
\usepackage{booktabs}

\DeclareMathOperator*{\argmin}{argmin}

\begin{document}
\title{On Altruism and Spite in Bimatrix Games}
\author{Michail Fasoulakis\inst{1}
\and
Leonidas Bakopoulos\inst{2}
\and
Charilaos Akasiadis\inst{2}
\and \\
Georgios Chalkiadakis\inst{2}
}
\authorrunning{Fasoulakis, Bakopoulos, Akasiadis and Chalkiadakis}
%
\institute{Royal Holloway, University of London\\
\email{michail.fasoulakis@rhul.ac.uk}
\and
Technical University of Crete\\
\email{\{lbakopoulos,cakasiadis,gchalkiadakis\}@tuc.gr}}
\maketitle           
\begin{abstract}
One 
common assumption in game theory is 
that any player optimizes 
a utility function that takes into account only its own payoff.
However, 
it has long been observed that in real life players may adopt an altruistic or even spiteful behaviour.
As such, there are numerous attempts in the economics literature 
that strive to explain 
the fact that players are not entirely selfish, but most of these works do not focus on the algorithmic implications of altruism or spite in games. 
In this paper, we relax 
the aforementioned ``self-interest'' 
assumption, and initiate the study of algorithmic aspects of bimatrix games---such as the complexity and the quality of their (approximate) Nash equilibria---under altruism or spite. We provide both a 
theoretical and an experimental treatment of these topics. Moreover, we demonstrate the potential for learning the degree of an opponent's altruistic/spiteful behaviour, and employing this 
for opponent selection and transfer of knowledge in bimatrix games.

\end{abstract}

\def\vzero{{\bm{0}}}
\def\vone{{\bm{1}}}
\def\vx{{\bm{x}}}
\def\vy{{\bm{y}}}
\def\ve{{\bm{e}}}

\section{Introduction}

One assumption widely used in game theory and economics, is the so-called Self-Interest Hypothesis (SIH), according to which players/agents\footnote{We will be using the terms {\em players} and {\em agents} interchangeably in this paper.} aim to maximize their personal payoff.
However, 
SIH validity in the real world has been questioned. Experiments 
in the 1980s and 1990s 
have demonstrated that people are not, in general, self-interested. As a result, several models attempt to explain this phenomenon by introducing concepts such as reciprocity, fairness, and altruism into decision-making processes~\cite{roth91,palfrey97,sobel2005,Fehr2006}.

As the seminal work of~\cite{sobel2005} points out, while 
several different models have been developed to describe and organize the evidence of non-selfish behaviour, there is no general model that provides a complete description of observational findings.
Several models put forward specific functional forms for {\em interdependent preferences}, assuming that individuals seek to
optimize
well-defined preferences which are nevertheless dependent on the behaviour of others; and then proceed to make predictions of agents' strategies based on the equilibrium behaviour ensuing from the utility functions describing the preferences in question. The notion of {\em spite}, which 
corresponds to various degrees of malicious or envious behaviour exhibited by individuals, also emerges in various manifestations given the different functional forms of preferences and utility defined in such models~\cite{levine1998,bolton00,sobel2005}.

In our work in this paper, we take a step towards 
an in-depth algorithmic study of economic models departing from the SIH hypothesis. We focus on bimatrix games, one of the most fundamental classes of games that can potentially give us insights to the study of games with more than two players. 

In such games we examine 
(both from a theoretical and from an experimental point of view) the computational complexity of equilibria arising when selfishness is relaxed and the concepts of {\em altruism} and {\em spitefulness} (or {\em spite}) are introduced. 
Note that by ``selfish'' we refer to rational agents whose utility is a function of their own payoff alone, i.e., theirs is a ``typical'' utility function; while altruistic or spiteful agents essentially maximize a utility function that also incorporates others' utility.

More specifically, our contributions are as follows.
We adopt the concept of {\em altruism} from~\cite{MM08} as was introduced for $n$-player games, 
and proceed to define {\em spite} in two-player (bimatrix) games in a straightforward manner. 
We define 
a game
in which players may exhibit a behaviour that is taking into account the utility of the opponents---i.e, may exhibit a level of altruism or spite.
First, we prove that in almost any case of altruism/spite, the computation of an exact NE remains PPAD-complete, and then we study approximate Nash equilibria (NE) under different levels of altruism/spite, providing 
polynomial-time algorithms for
specific fundamental cases. 

Interestingly, for specific levels of altruism or spite, not only can we find {\em in polynomial-time} an $\varepsilon$-NE that is {\em better} than the state-of-the-art approximation for polynomial-time algorithms for approximate Nash equilibria \cite{DFM23}, but we also show this has {\em maximum social welfare}.

In addition, we design an algorithmic framework and test altruism/spite in numerous bimatrix games. 
We put forward a gradient descent-based algorithm that identifies strategies along with an optimal setting of the spite/altruism parameters, such that an $\varepsilon$-equilibrium with low $\varepsilon$ is found. 

We test our algorithm experimentally for computing 
an
 approximate NE in games considering altruism/spite.

Moreover, we provide experiments to showcase
the
{\em learning} 
of
the $\lambda$ values corresponding to the 
{\em unknown}
{\em levels of altruism/spite} of opponents;
and
thus
allow rational agents to use 
learned values to beneficially ``select'' which opponent to face in a bimatrix game.
In addition, we demonstrate the potential for the transfer of knowledge regarding the opponents' levels of altruism/spite, and for ``transfer learning''~\cite{pan2010} among different games.
As such, our work takes a first step in combining algorithmic game theory with machine learning within this framework. 

\section{Preliminaries}

\paragraph{Basic game-theoretic concepts.}
We consider bimatrix games\footnote{Any bimatrix game can be transformed to a bimatrix game of entries in $[0,1]$ with exactly the same Nash equilibria.} $G=(R,C) \in [0,1]^{n \times n}$, where $R$ is the payoff matrix of the row player (the ``Row'') and $C$ the payoff matrix of the column player (the ``Column''). Any player has $n$ pure strategies at her disposal and any pure strategy $i$ is denoted as the column vector $\ve_i$, which has $1$ in the index $i$ and zero elsewhere.
A player can also play mixed strategies, as probability distributions $\vx \in \Delta^{n-1}$ on her pure strategies, where $\Delta^{n-1}$ is the $n-1$ dimensional simplex created by her pure strategies. A pair of mixed strategies $(\vx,\vy)$ is called a strategy profile, where $\vx,\vy$ are the mixed strategies of the 
Row and Column, respectively.

Let $(\vx,\vy)$ be a strategy profile, then the payoff for the Row is $\vx^TR\vy$ and the payoff for the Column is $\vx^TC\vy$. In this paper, we consider that the players are expected payoff maximizers, in other words, they play strategies to maximize their expected payoff, given the strategy of the other player.
A strategy profile is a social optimum in the utilitarian sense (sum of the payoffs) if it is the maximum over all strategy profiles. We can easily see that in bimatrix games there is at least one pure strategy profile that is a social optimum, and that it can be computed in polynomial-time by exhaustive search over the pure strategy profiles.

For any strategy profile $(\vx,\vy)$ we define as {\em regret} or {\em approximation} of this strategy profile for the Row player
the function:
\begin{equation*}
\begin{split}
f_R(\vx,\vy): \Delta^{n-1} \times \Delta^{n-1} \to [0,1], \text{ with } \\
f_R(\vx,\vy) =  \max_i \ve^T_iR\vy-\vx^TR\vy,
\end{split}
\end{equation*}
and, similarly, for the Column the function:
\begin{equation*}
\begin{split}
f_C(\vx,\vy): \Delta^{n-1} \times \Delta^{n-1} \to [0,1], \text{ with } \\
f_C(\vx,\vy) = \max_j \vx^TC\ve_j - \vx^TC\vy.
\end{split}
\end{equation*}

 \noindent The (approximate) Nash equilibria are defined as follows.

\begin{definition}[Nash equilibria (NE)]
A strategy $(\vx,\vy)$ is a Nash equilibrium, if and only if, for any $i,j$, it holds that:
\begin{equation*}
\begin{split}
\vx^TR\vy \geq \ve^T_iR\vy, \text{ and }
\vx^TC\vy \geq \vx^TC\ve_j,    
\end{split}
\end{equation*}
\end{definition}
\noindent or, in other words $f_R(\vx,\vy) = 0$ and $f_C(\vx,\vy) = 0$ (the regrets of both players are zero).
Furthermore, for approximate Nash equilibria, we have the following definition.

\begin{definition}[$\varepsilon$-approximate Nash equilibria]\label{def:eps_nash}
A strategy $(\vx,\vy)$ is an $\varepsilon$-approximate Nash equilibrium, for any $\varepsilon \in [0,1]$, if and only if, for any $i,j$, it holds that:
\begin{equation*}
\begin{split}
\vx^TR\vy +\varepsilon\geq \ve^T_iR\vy, \text{ and }
\vx^TC\vy +\varepsilon \geq \vx^TC\ve_j,
\end{split}
\end{equation*}
\end{definition}
\noindent or in other words $f_R(\vx,\vy)\leq \varepsilon$ and $f_C(\vx,\vy) \leq \varepsilon$ (the regrets of both players are at most $\varepsilon$). It is easy to see that we have an exact NE if and only if $\varepsilon = 0$.

\paragraph{Altruism and spite.}

In this paper, we consider the level of altruism or spite 
that players introduce as a policy/behaviour to an initial game $(R,C)$, with two parameters $\lambda_R, \lambda_C \in [-1,1]$, for the row and Column, respectively. 
The game is thus
transformed 
to the following modified bimatrix
 game:
\begin{equation*}
\begin{split}
{G'} = ({R'},{C'}) = \Big(R+\lambda_R\cdot C, C + \lambda_C \cdot R \Big).
\end{split}
\end{equation*}

In particular, we consider the level of altruism for strictly positive values of $\lambda$'s as
in ~\cite{MM08}. 
\begin{definition}[Altruism ~\cite{MM08}]
The Row is altruistic of a level $\lambda_R$, if and only if, $\lambda_R>0$.
Similar for the Column.
\end{definition}

In the same spirit, here we introduce and define the level of spite (or spitefulness) for strictly negative values.
\begin{definition}[Spite]
The Row is spiteful of a level $|\lambda_R|$, if and only if, $\lambda_R<0$.
Similar for the Column.
\end{definition}

The remaining case of $\lambda$'s being $0$, is the case where ${G'} = ({R'},{C'}) = (R, C) = G$, i.e., the 
initial and the modified games
coincide.

Note that an agent's strategy depends on the policy/behavior (altruistic, spiteful, or SIH-compliant) that is
``prescribed'' to her by her $\lambda$,
which drives her to maximize
her expected payoff in the modified game $G'$.

\section{Game theoretic aspects of altruism and spite}
We now proceed to make some initial observations for the modified payoff game in a game theoretic context.
In the case of altruism the payoff matrix of the Row is:
\begin{equation*}
\begin{split}
{R'} = R + \lambda_R \cdot C = (1-\lambda_R)\cdot R+\lambda_R\cdot (C+R),
\end{split}
\end{equation*}
similar for the Column.
In this case,
the payoff matrix 
is a convex combination between the own payoff matrix (selfish interest) and the social optimum (altruistic interest), where the latter is the utilitarian optimum (the sum of payoffs).

In the case of {\em spite} the Row's payoff matrix  is:
\begin{equation*}
\begin{split}
{R'} = (1+\lambda_R)\cdot R-\lambda_R\cdot (R-C).
\end{split}
\end{equation*}
Interestingly, this is a convex combination\footnote{Note that $\lambda_R$ is a negative number.} between the own payoff matrix (selfish interest) and a matrix with payoff $R-C$ (spiteful interest), where the Row maximizes the difference of the payoffs between her own payoff and the payoff of the other player. Similar for the Column.

Another interesting observation is that in the case of equivalent policy/behaviour (both same level of altruism, spite, or selfish), in other words $\lambda = \lambda_R = \lambda_C$, we have: 
\begin{align*}
&{G'} = \Big(R+\lambda \cdot C, C+\lambda \cdot R \Big)\\
&= (1-\lambda) \cdot \frac{(R-C,C-R)}{2} 
+ (1+\lambda) \cdot \frac{(R+C,R+C)}{2}.
\end{align*}

We see that if $\lambda = -1$ the game is zero-sum, and if $\lambda = 1$ 
it 
is a potential game/common payoff---both poly-time solvable. In any other case, we have a general-sum game. This interesting result can be seen as a complement to the well-known result that any bimatrix game (i.e., one with $\lambda = 0$ in our context) is the average of one zero-sum and one potential game \cite{HR20}.
Finally, if 
Row
is altruistic
with
$\lambda>0$, and  Column 
spiteful 
by the same level $\lambda$ (or symmetrically), then 
$G'$ 
becomes:
\begin{align*}
&{G'} = \Big(R+\lambda \cdot C, C-\lambda \cdot R \Big)\\
&= (1-\lambda) \cdot (R, C) 
+ \lambda \cdot (R+C,C-R).
\end{align*}

In this case, we see that the modified game is a convex combination of the initial selfish interest one, and the game where one player maximizes the social optimum, while the other the difference between hers 
and the 
opponent's payoff.

\section{Computational complexity of computing a Nash Equilibrium under altruism or spite}

We know that computing a NE in bimatrix games is PPAD-complete \cite{CDT09,DGP09} when both players seek to maximize their own payoff only. 
On the other hand, if players have an altruistic or spiteful behaviour, the Nash equilibria of the modified game may change and in some cases be completely different.
Thus, we are interested whether this change can affect the computational complexity of the problem.
We study this under the presence of different combinations and levels of altruistic/spiteful players' behaviours.

It is easy to see that in the extreme case of $\lambda_R = \lambda_C = 1$ (both players have the maximum level of altruism), a strategy profile that maximizes the utilitarian optimum (sum of payoffs) of the game $G$ is an NE in the game ${G'}$. This strategy profile can be easily found in polynomial time by exhaustive search on the pure strategy profiles of the game.
Furthermore, in the other extreme case of $\lambda_R = \lambda_C = -1$ (both players have the maximum level of spite), the game ${G'}$ is a zero-sum game, so we can compute again in polynomial time a NE by linear programming \cite{Adler13}. For almost any remaining case\footnote{We leave open the case of $\lambda_R \cdot \lambda_C = -1$.}, we have the following Theorem.

\begin{theorem}
\label{Complexity theorem}
For any $\lambda_R\in (-1,1)$ and any $\lambda_C\in (-1,1)$, the problem of computing a NE in  
${G'}$ is PPAD-complete. 
\end{theorem}
\begin{proof}
Let an arbitrary bimatrix game $(A,B) \in [0,1]^{n \times n}$.
As mentioned above, by \cite{CDT09}, we know that the problem of computing a NE in a bimatrix game is PPAD-complete.
Let now the game $$(R,C) = (A-\lambda_R \cdot B, B - \lambda_C \cdot A),$$ then we have that 
\begin{align*}
&{G'} = \Big(A-\lambda_R \cdot B + \lambda_R \cdot B - \lambda_R \cdot \lambda_C \cdot A, \\ 
& ~~~~~~~~~~~~ \quad  B - \lambda_C \cdot A + \lambda_C \cdot A - \lambda_C \cdot \lambda_R \cdot B\Big)\\
&= \Big((1-\lambda_R \cdot \lambda_C) \cdot A , (1-\lambda_R \cdot \lambda_C) \cdot B \Big),\\
& = (1-\lambda_R \cdot \lambda_C) \cdot (A,B),
\end{align*}
which we can easily prove, since $(1-\lambda_R \cdot \lambda_C)>0$, that any NE of this game is also a NE in the game $(A,B)$.
\end{proof}

Note that the previous result does not imply anything about the complexity of the initial game $(R,C)$ from which the reduction starts. 
It does imply, however, that in general if we introduce altruism/spite, the computation of a Nash equilibrium in the modified game is PPAD-complete.

Having these negatives results, assuming that P $\neq$ PPAD, we are interested in the effect of altruism/spite in the computation of approximate Nash equilibria in our context as we see in the following section.

\section{Regret under altruism or spite}

With the previous complexity results for the exact Nash equilibria at hand, we now focus on the case of approximate NE in our context. 
First, we study some fundamental cases where the players can indeed achieve approximate Nash equilibria in polynomial-time with low regret under altruism/spite, and then we give an algorithmic framework as a ``learning'' algorithm where the players in an alternating optimization fashion can compute their behavior/policy in order to achieve approximate Nash equilibria with good regret.
Our initial results are the following.

\begin{theorem}
Let a bimatrix game $G = (R,C) \in [0,1]^{n \times n}$ and the modified game $G' = (R+\lambda \cdot C, C+\lambda \cdot R) \in[0,1]^{n \times n}$, with $\lambda >0$ (Both players are altruistic of the same level). Then,  
the social optimum (utilitarian optimum) of the game $G'$ is an $(1-\lambda )$-Nash equilibrium.
\end{theorem}
\begin{proof}
Let $(e_i,e_j)$ be the social optimum (Utilitarian optimum) of the game $G'$. Then, it holds that, for any $k,\ell$:
\begin{equation*}
\begin{split}
(1+\lambda) \cdot R_{ij} + (1+\lambda) \cdot C_{ij} &\\
\geq (1&+\lambda) \cdot R_{k \ell} + (1+\lambda) \cdot C_{k \ell}.
\end{split}
\end{equation*}
Consider now the case that $\ell = j$.
Since, $1+\lambda>0$, the previous inequality implies that 
\begin{equation*}
\label{SO}
\begin{split}
R_{ij} + C_{ij} \geq R_{k j}+C_{kj},\\
\rightarrow\\
R_{ij} + C_{ij} -(1-\lambda) \cdot C_{ij} -(1-\lambda) \cdot C_{kj} \geq R_{k j}+C_{kj}\\-(1-\lambda) \cdot C_{ij}-(1-\lambda) \cdot C_{kj},\\
\rightarrow\\
R_{ij} 
+\lambda \cdot C_{ij} -(1-\lambda) \cdot C_{kj} \geq R_{k j}+\lambda \cdot C_{kj}\\-(1-\lambda) \cdot C_{ij},\\
\rightarrow\\
R_{ij} 
+\lambda \cdot C_{ij} +(1-\lambda) \cdot (C_{ij}-C_{kj})\\ \geq R_{k j}+\lambda \cdot C_{k j}.\\
\end{split}
\end{equation*}
Since $C \in [0,1]^{n \times n}$, then $C_{ij} -C_{kj} \leq 1$. This, by definition, implies that $(e_i,e_j)$ has regret at most $(1-\lambda)$. Similar, for the other player.
\end{proof}
Note that in the case of study of the game $G'$ in the previous Theorem, the Utilitarian optimum of the game $G'$ is also an optimum for the game $G$. Thus, our Theorem implies that we can in poly-time find a strategy profile with $(1-\lambda)$ approximation and also maximum social welfare, even for the initial game. Furthermore, in the case of level of altruism $\lambda > \frac{2}{3}$, not only we can find in polynomial-time an $\varepsilon$-NE with $\varepsilon < \frac{1}{3}$ that is better than the state of art approximation for polynomial-time algorithms for approximate Nash equilibria \cite{DFM23}, but also this has maximum social welfare in the game. We now give a similar result for the case of spite as follows: 

\begin{theorem}
Let a bimatrix game $G = (R,C) \in [0,1]^{n \times n}$ and the modified game $G' = (R+\lambda \cdot C, C+\lambda \cdot R) \in[0,1]^{n \times n}$, with $\lambda <0$ (Both players are spiteful of the same level). Then, there is a polynomial time to compute an $(1+\lambda)$-Nash equilibrium.
\end{theorem}
\begin{proof}
We can analyse the game $G'$ as $G' = (R-C, C-R)+(1+\lambda)\cdot (C,R)$.
For the zero-sum game $(R-C,C-R)$ we can find in polynomial time a Nash equilibrium $(x,y)$ by Linear programming \cite{Adler13}. Then, it holds by the definition of the NE that $x^T(R-C)y \geq e^T_i(R-C) y$, for any $i$. Thus, we have that:
\begin{equation*}
\begin{split}
x^T(R+\lambda \cdot C)y = x^T(R-C)y + (1+\lambda) \cdot  x^TCy \\
\geq e^T_i(R-C) y + (1+\lambda) \cdot x^TCy,\\
\rightarrow\\
x^T(R+\lambda \cdot C)y + e^T_iCy 
- (1+\lambda) \cdot  x^TCy
\geq e^T_iRy,\\
\rightarrow\\
x^T(R+\lambda \cdot C)y + e^T_iCy + \lambda \cdot e^T_iCy - (1+\lambda) \cdot x^TCy\\
\geq e^T_iRy + \lambda \cdot e^T_iCy,\\
\rightarrow\\
x^T(R+\lambda \cdot C)y +  (1+\lambda) \cdot e^T_iCy
\geq e^T_i(R + \lambda \cdot C)y.\\
\end{split}
\end{equation*}
The last inequality holds since $(1+\lambda) \cdot x^TCy \geq 0$.
Thus, we have that $(1+\lambda) \cdot e^T_iCy \leq 1+\lambda$, since $e^T_iCy \leq 1$, and this implies that the regret of the row player is at most $(1+\lambda)$. Similar for the column player.
\end{proof}

Similarly as above, we can see that in the case of level of spite of $\lambda < -\frac{2}{3}$ (high spite), we can find in polynomial-time an $\varepsilon$-NE, with $\varepsilon < \frac{1}{3}$, better than the state of art approximation algorithms of Nash equilibria in bimatrix games \cite{DFM23}.

For other cases of altruism/spite we will introduce an algorithmic framework of study not only the regret, but also the optimal policies of the player to converge to approximate NE with good regret. This framework is based on the idea of the descent method of the paper of \cite{TS08} to achieve approximate NE in bimatrix games.

This algorithm is based on performing gradient descent to minimize the maximum of the two regrets function (we will refer to this as {\em the TS function} for now), a method proposed in~\cite{TS08} for computing a $0.3393$-NE. The high level idea is that the descent method converges to a stationary point, and then, given its properties, a more sophisticated strategy profile can be constructed, so that this profile or the stationary one have no more than $0.3393$ regret.

Here, we only focus on the descent method and the regrets of the stationary point that are reached.
But before we introduce our framework, we give some simple illustrative examples that show the possible change of the stationary points under altruism/spite that inspired our approach.

As proven in~\cite{CDHLL23}, there is a tight example/game with a stationary point of the TS function of approximation 0.3393. Inspired by this, we provide and study the following similar example as:
\begin{equation}
\label{game:ts_problem}
R = \begin{pmatrix}
0.1 & 0 & 0 \\
0.1 + b & 1 & 1 \\
0.1 + b & 0 & 0
\end{pmatrix},
C = \begin{pmatrix}
0.1 & 0.1 + b & 0.1 + b \\
0 & 1 & 0 \\
0 & 1 & 0
\end{pmatrix}
\end{equation}
with $b$ equal to 0.3393. 
Note that in this game the strategy profile $x_s = y_s = [1 \quad 0 \quad 0 ]$ is a stationary point with $0.3393$ approximation and the point $x^* = y^* = [0 \quad 1 \quad 0 ]$ is an exact Nash equilibrium. 
Let us now consider the case of introducing a level of altruism/spite with $\lambda$'s in this game.
The modified game ${G'} = ({R'},{C'})$ will be as follows:

\label{game:ts_problem2}
\begin{align*}
{R'} = \begin{pmatrix}
0.1 + \lambda_R \cdot 0.1 & \lambda_R \cdot (0.1+b) & \lambda_R \cdot (0.1+b) \\
0.1 + b & 1+ \lambda_R & 1 \\
0.1 + b & \lambda_R & 0
\end{pmatrix}
,\\
\quad
{C'} = \begin{pmatrix}
0.1 + \lambda_C \cdot 0.1 & 0.1 + b & 0.1 + b \\
\lambda_C \cdot (0.1+b) &1 +\lambda_C &  \lambda_C\\
\lambda_C \cdot (0.1+b) & 1 & 0
\end{pmatrix}.
\end{align*}

When $\lambda_R =0$ and $\lambda_C=\frac{1}{2}$, 
the modified game\footnote{Here for simplicity of the presentation we don't scale the payoffs of the players to $[0,1]$.} becomes
\begin{equation*}
\label{game:ts_problem3}
{R'} = \begin{pmatrix}
0.1 & 0 & 0 \\
0.1 + b & 1 & 1 \\
0.1 + b & 0 & 0
\end{pmatrix},
{C'} = \begin{pmatrix}
0.15 & 0.1 + b & 0.1 + b \\
\frac{0.1+b}{2} &1.5&  \frac{1}{2}\\
\frac{0.1+b}{2} & 1 & 0
\end{pmatrix}.
\end{equation*}
We can see that, since  the regrets in the modified game at $(\vx_s,\vy_s)$ are not anymore equal, this strategy profile cannot be a stationary point in the new game for the TS objective function, since it has been proven that any stationary point must have the same regrets (see e.g. Lemma 2.4 in ~\cite{F17}).
This example shows us that the stationary points can be changed under altruism or spite.

However, 
this is not always the case. 
E.g.,
consider the following game: 
\begin{equation}
\label{game:ts_problem4}
R = \begin{pmatrix}
0 & 0 & 0 \\
b & 1 & 1 \\
b & 0 & 0
\end{pmatrix},
\quad
C = \begin{pmatrix}
0 & b & b \\
0 & 1 & 0 \\
0 & 1 & 0
\end{pmatrix},
\end{equation}
In this game 
$x_s = y_s = [1 \quad 0 \quad 0]$ is also a stationary point.
Let now the modified game with $\lambda_R =0 ,\lambda_C = 1/2$, as
\begin{equation*}
\label{game:ts_problem5}
{R'} = \begin{pmatrix}
0 & \lambda_R \cdot b & \lambda_R \cdot b \\
b & 1+ \lambda_R & 1 \\
b & \lambda_R & 0
\end{pmatrix},
{C'} = \begin{pmatrix}
0 &  b &  b \\
\lambda_C \cdot b &1 +\lambda_C &  \lambda_C\\
\lambda_C \cdot b & 1 & 0
\end{pmatrix}.
\end{equation*}
Then, 
$\vx_{s}=\vy_{s}=[1 \quad 0 \quad 0]$ remains a stationary point. Inspired by these observations, we provide an algorithmic framework based on gradient-based approaches to achieve optimal behavior/policies for better approximation of the Nash equilibria for the strategy profile of the players. 
\paragraph{{\bf Our Algorithmic framework.}}
Motivated by the example above, we now propose an algorithmic framework in which players start the game with a ``selfish'' behaviour ($\lambda$'s $=0$), and subsequently change it---or, in other words, {\em learn a different policy/ behaviour}---so as to achieve {\em a better}, in terms of regret in the modified game, strategy profile.

In particular, 
Algorithm~\ref{prob:ts_problem} constitutes an alternating process that optimizes, in terms of regret, the strategy profiles $(\vx^t,\vy^t)$ given the level of altruism/spite $(\lambda^{t-1}_R,\lambda^{t-1}_C)$, and vice versa.
Specifically,
starting from $(\lambda^0_R,\lambda^0_C) = (0,0)$, we apply the projected gradient descent~\cite{bertsekas1997nonlinear} (PGD) procedure on a quadratic program
with fixed values of $\lambda^{t-1}$'s, having 
a particular
objective
regarding 
the regrets, for instance minimizing the maximum of the regrets as indicated in program (\ref{prob:quad}) below,\footnote{Alternatively,
the sum of the regrets, as in program~\ref{prob:quad2}.}
to converge to stationary point strategies $(\vx^t,\vy^t)$. Then, having as fixed the strategy profile $(\vx^t,\vy^t)$, the stationary point of the previous step, we find the optimal policies $(\lambda^{t}_R,\lambda^{t}_C)$, that minimizes the objective function, by solving a linear program, LP (\ref{prob:lin}). 
This procedure is repeated for a maximum of $h$ times, or as the algorithm converges.

\begin{algorithm}
\caption{An alternating optimization process to produce a regret-minimizing  $\langle(\vx^t,\vy^t),(\lambda^t_R, \lambda^t_C) \rangle$ pair}
\label{prob:ts_problem}
\begin{algorithmic}[1]
\STATE \textbf{Input:} A bimatrix game $(R,C) \in [0,1]^{n \times n}$, an initial strategy profile $(\vx^0, \vy^0)$ and $(\lambda^0_R,\lambda^0_C)$.
\FOR{$t = 1$ to $h$}
    \STATE \textit{// Find a stationary point $(\vx^t,\vy^t)$ given $(\lambda^{t-1}_R, \lambda^{t-1}_C)$ }
    \STATE $(\vx^t,\vy^t) \leftarrow$ Solution of the quadratic program (\ref{prob:quad}) using PGD with $\lambda_R = \lambda^{t-1}_R$ and $\lambda_C = \lambda^{t-1}_C$.
    \STATE \textit{// Find optimal $(\lambda^t_R, \lambda^t_C)$ given strat. profile $(\vx^t,\vy^t)$}
    \STATE $(\lambda^t_R, \lambda^t_C) \leftarrow$ Solution of the linear program (\ref{prob:lin}) with $\vx = \vx^t$ and $\vy = \vy^t$.
\ENDFOR
\STATE \textbf{Return} $(\vx^{t-1},\vy^{t-1})$ and $(\lambda^{t-1}_R, \lambda^{t-1}_C)$
\end{algorithmic}
\end{algorithm}

In more detail, 
with respect to the first part of the strategy profiles optimization given $\lambda$'s, we adopt a quadratic formulation of (\ref{prob:quad}) inspired by 
\cite{MS64,TS08,DFM23}:

\begin{align}
\begin{split}
\min_{\vx,\vy, \pi_1,\pi_2} \max&\Big\{ \pi_1 - \vx^T R \ \vy - \lambda_R \ \vx^TC \ \vy,
 \\& \quad \quad \pi_2  - \vx^T C \ \vy -  \lambda_C \ \vx^T R \ \vy \Big\} \\
\text{s.t.} \quad
\ve^T_i (R&+\lambda_R \ C) \ \vy \leq \pi_1, \text{ for any }i \in [n],  \\
 \quad \vx^T (C&+\lambda_C \ R) \  \ve_j \leq \pi_2,
\text{ for any } j \in [n], \\
 \quad \sum_i \vx_i &=1;
 \sum_j \vy_j =1;
 \pi_1,\pi_2 \in [-1,1].
\end{split}
\label{prob:quad}
\end{align}

Note that in \cite{MS64} a similar quadratic formulation was proposed:

\noindent

{
\begin{align}
\begin{split}
\min_{\vx,\vy, \pi_1,\pi_2}&  \pi_1 - \vx^T R \ \vy - \lambda_R \ \vx^T C \ \vy\\
\quad \quad \quad \quad &\quad \quad \quad \quad + \pi_2  - \vx^T C \ \vy-  \lambda_C \ \vx^T R \ \vy \\ 
\text{subject to} \quad
&\ve^T_i (R+\lambda_R \ C) \  \vy \leq \pi_1, \text{ for any }i \in [n],\\
& \vx^T (C+\lambda_C \ R) \ \ve_j \leq \pi_2,
\text{ for any } j \in [n],\\
& \sum_i \vx_i =1;
\sum_j \vy_j =1; 
\pi_1,\pi_2 \in [-1,1].
\end{split}
\label{prob:quad2}
\end{align}
}
In the above quadratic programs, given altruism/spite levels $(\lambda_R,\lambda_C)$ of the players, the result $(\vx,\vy)$ is a stationary point for the modified game.
We note here that the only exact Nash equilibria of the game are the global optima, but the stationary points that are different than the global optima are nevertheless $\varepsilon$-approximate Nash equilibria, with $\varepsilon>0$. It is well-known that such a formulation can lead to a stationary point that is not necessarily a Nash equilibrium.

On the other hand, given a strategy profile we solve the following linear program (\ref{prob:lin}) for finding the optimal behaviour for minimizing the maximum of the regrets:
\begin{equation}
\begin{split}
\min_{(\lambda_R, \lambda_C)} & \gamma \\
\text{s.t.} \quad
&\gamma \geq \ve^T_i(R+\lambda_R \ C)\ \vy -  \vx^T(R+\lambda_R \ C)\ \vy, \\
& \gamma \geq \vx^T (C+\lambda_C \ R)\ \ve_j - \vx^T(C+\lambda_C \ R) \ \vy, \\
& \quad \quad \quad \quad \quad  \quad  \text{for any } i,j\in [n]; \lambda_R, \lambda_C \in [-1,1]. 
\end{split}
\label{prob:lin}
\end{equation}
The following is the LP finding the optimal behavior for minimizing the sum of the regrets function:
\begin{align}
\begin{split}
\min_{(\lambda_R, \lambda_C)} & \gamma \\
\text{s.t.} \quad
&\gamma \geq \ve^T_i(R+\lambda_R \ C)\ \vy -  \vx^T(R+\lambda_R \ C) \  \vy   \\
& + \vx^T  (C+\lambda_C \ R) \ \ve_j - \vx^T (C+\lambda_C \ R)\ \vy, \\
& \quad \quad \quad \quad \quad \quad  \text{for any } i,j\in [n]; \lambda_R, \lambda_C \in [-1,1]. 
\end{split}
\label{prob:lin2}
\end{align}

\paragraph{{\bf Analysis of  Algorithm \ref{prob:ts_problem}}.}

Theorem~\ref{thm:alg_converges} shows the convergence of Algorithm~\ref{prob:ts_problem}:

\begin{theorem}
\label{thm:alg_converges}
Let a bimatrix game $G = (R,C) \in [0,1]^{n \times n}$ and $r^t$ be the regret function\footnote{Maximum regret, or sum of the regrets.} of the two players in the modified game $G' = (R + \lambda^t_R \cdot C,C+\lambda^t_C \cdot R) \in [0,1]^{n \times n}$, for any $t\geq 0$.
Then, it holds that $r^{t+1} < r^{t}$, for any $t$ s.t. $(\vx^t,\vy^t)$ is not a stationary point given $(\lambda^t_R,\lambda^t_C)$, and $(\lambda^t_R,\lambda^t_C)$ is not an optimal pair given $(\vx^t,\vy^t)$.
\end{theorem}
\begin{proof}
Let an iteration $t$ of the algorithm, with $(\vx^{t},y^{t})$ be a strategy profile, $(\lambda^t_R,\lambda^t_C)$ be the current a pair of $\lambda$'s and $r^t$ be the value if the regret function in the modified game in this strategy profile. The PGD algorithm will find a new stationary point $(\vx_s,\vy_s)$, minimizing the maximum of the regrets, s.t. $r_s < r^t$ by definition of the quadratic programming (3), since $(\vx^t,\vy^t)$ is not optimal given $(\lambda^t_R,\lambda^t_C)$. Let now that we fix this profile, then the LP (5), by definition, will find a new pair of $\lambda$'s, $(\lambda^{t+1},\lambda^{t+1}_C)$, s.t. the new maximum regret $r^{t+1}$ will be less or equal to the previous $r_s < r^t$. This holds for any iteration $t$, and since the regret is bound by below by zero, Alg. 1 will converge.
\end{proof}
Theorem~\ref{thm:alg_converges} 
straightforwardly implies the
following:

\begin{corollary}
Let a bimatrix game $G = (R,C) \in [0,1]^{n \times n}$. Then, our algorithm will converge to a stationary point of the 
game $G' = (R + \lambda^{*}_R \cdot C,C + \lambda^{*}_C \cdot R)$, with $(\lambda^{*}_R,\lambda^{*}_C)$ being the final assigned $\lambda$'s.
\end{corollary}

Note the previous results imply that the reached strategy profile $(\vx^{*},\vy^{*})$ is a stationary point 
given $(\lambda^{*}_R,\lambda^{*}_C)$, and the pair $(\lambda^{*}_R,\lambda^{*}_C)$ is an optimal behaviour given $(\vx^{*},\vy^{*})$.

\section{Experimental Evaluation}
Here, we first examine experimentally the effects of altruistic/spiteful behaviour on the regret of the modified game.
We then show 
the potential for
learning the degree of altruistic/spiteful behaviour of potential opponents, and employ the knowledge acquired to identify the most ``preferred'' opponent. 
Moreover, we showcase the potential for the {\em transfer of knowledge} 
regarding an opponent's $\lambda$
and for {\em transfer learning} 
between different game settings.\footnote{The experiments were run on an Ubuntu 22.04 computer.
This computer has a Ryzen 7, 16-core CPU and 16GB of
ram.}

\subsection{Approximate NE under altruism and spite}

Here, we present experiments to assess the value of approximate NE under various levels of altruism, and to test Alg.~\ref{prob:ts_problem}.

In our first experiments, we will compute approximate NE under various $(\lambda_R, \lambda_C)$ values, using PGD to solve either Problem~\ref{prob:quad}, or Problem~\ref{prob:quad2} for different  $\lambda_R, \lambda_C$ values. Following that, we will characterize an approximate NE considering its $\varepsilon$ value.

To begin, for each bimatrix game, we create $21\times21 = 441$ modified ones; each modified game has a different combination of $\lambda$ values for the two players. The $441$ games were derived by combining $21$ different values for $\lambda$ for each player. These correspond to numbers ranging from $-1$ to $1$ in increments of $0.1$ (e.g., $[-1.0, -0.9, -0.8, \ldots, 0.8, 0.9, 1.0]$). For each game, we constructed a $21 \times 21$ mesh grid that plots the $\varepsilon$ approximation for each combination of $\lambda$ values. We calculate $\varepsilon$
after PGD converged to a strategy profile $(\vx, 
\vy )$ for both players. For calculating the strategy profile, we solve (via PGD) Problems~\ref{prob:quad2} (or Problems~\ref{prob:lin2}).

Note that we address the dependence of the optimization problem's solution on the initial point by solving each problem multiple (specifically, 20) times, 
each with a distinct initial strategy profile $(\vx,\vy)$.  Unless otherwise specified, initial strategy profiles were chosen randomly by sampling  $2n$ samples (where $n > 1$ is the number of distinct actions) from a standard normal distribution, and applying softmax in order to ensure $\sum_{i=1}^nx_i = \sum_{i=1}^n y_i = 1$ . Our final results
are then computed as averages over those runs.

The bimatrix game (\ref{game:penalty_game_problem}) below describes the payoff matrices for the {\em Penalty Game (PG)}
in~\cite{ClausBoutilier1998}. 
Parameter $k$ is usually set to a high negative value to express a high penalty. In our experiment we set $k=-100$. This game has three pure NE points, $(x_i, x_i)$, for $i=1,2,3$.
However, 
those equilibrium points are not equivalent:
e.g., the $(x_2,y_2)$ strategy leads to a pure NE with $2$ for each player, while $(x_1, y_1)$ rewards each player with $10$.  

\begin{equation}
\label{game:penalty_game_problem}
R = C = \begin{pmatrix}
10 & 0 & k \\
0 & 2 & 0 \\
k & 0 & 10
\end{pmatrix}.
\end{equation}

\begin{figure}[h!]
    \centering
    \includegraphics[width=0.55\linewidth]{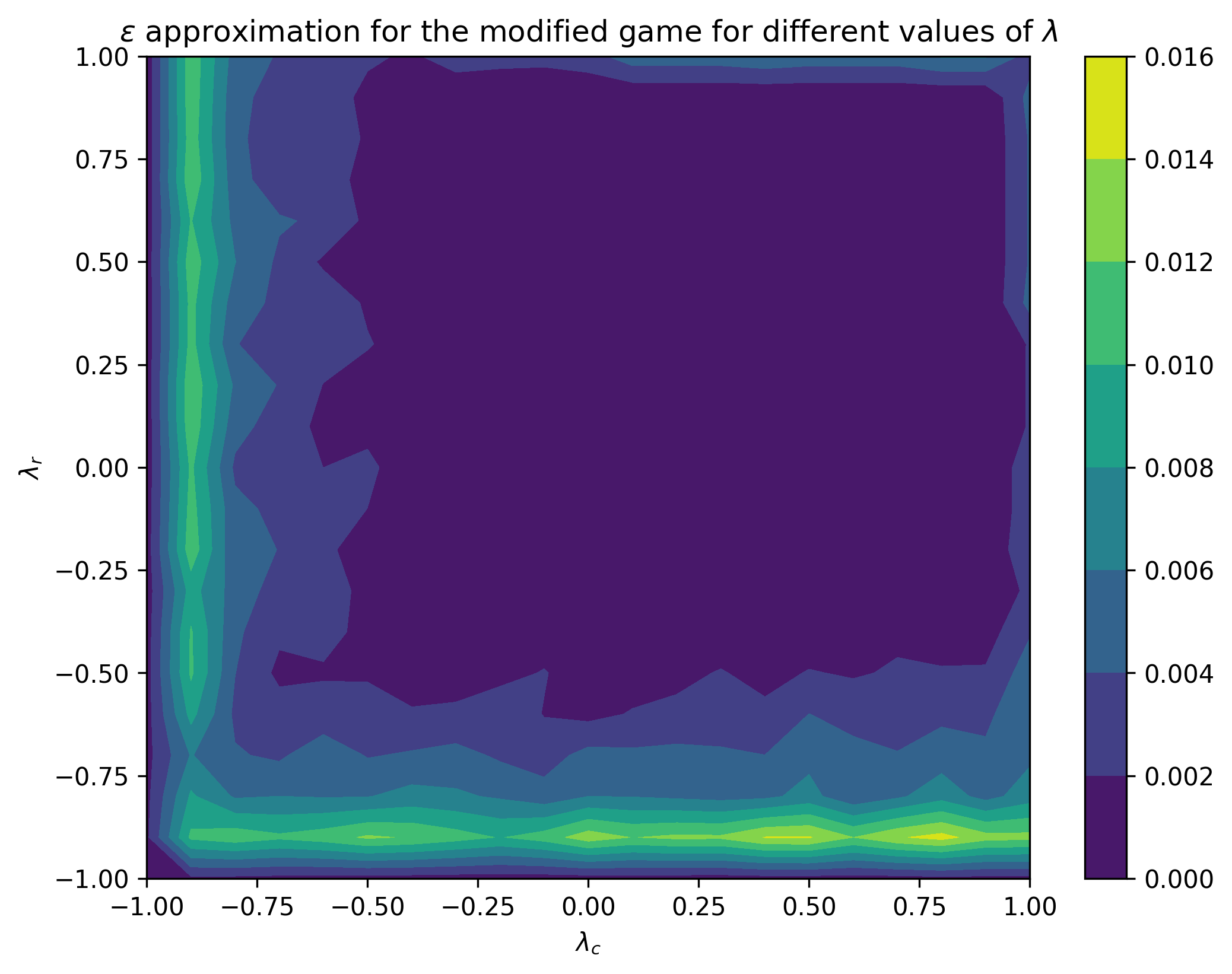}
    \caption{The $\varepsilon$ approximation for PG for the modified game. Results are averages over 20 runs.}
    \label{fig:Penalty_Game_approximations}
\end{figure}

Figure~\ref{fig:Penalty_Game_approximations} reports the approximation of NE for different values of $\lambda$,  calculated as in Definition~\ref{def:eps_nash}, using the strategy profile after convergence. As seen in Figure~\ref{fig:Penalty_Game_approximations}, the addition of altruism/spite does not drastically change the already low (near-zero) approximation. In other words, PGD converges to a low approximation $\varepsilon$-NE for most combinations of $\lambda$ values. However, when the behaviour of at least one player is described by a $\lambda \approx -0.8$, then the approximation slightly increases.

\begin{figure}[h!]
    \centering
    \includegraphics[width=0.55\linewidth]{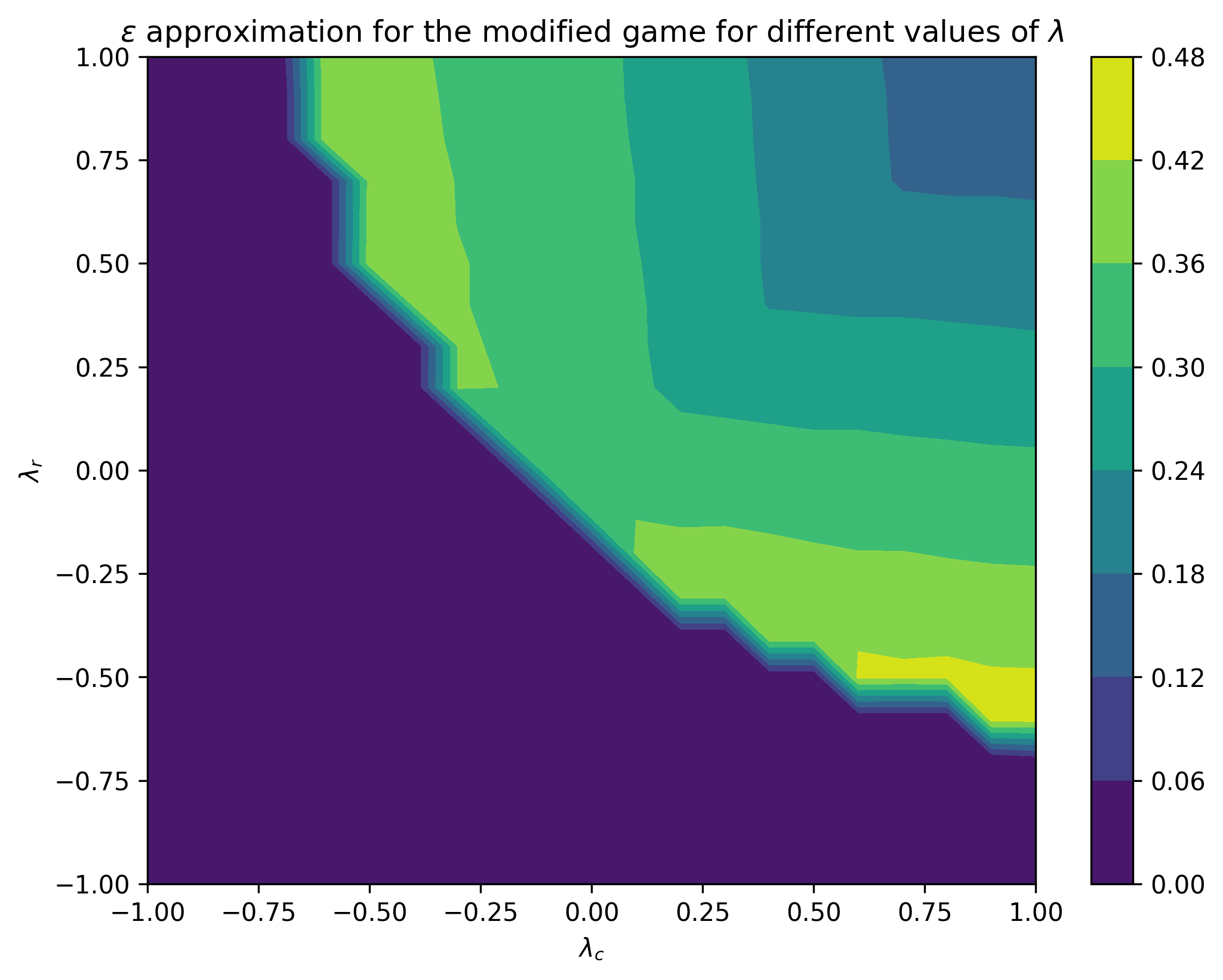}
    \caption{Approximation for the TS tight example \cite{CDHLL23} for the modified game.}
    \label{fig:TS_approximations}
\end{figure}

Figure~\ref{fig:TS_approximations} shows the approximations in the TS tight example \cite{CDHLL23} after the agents converged to a strategy profile solving Program~\ref{prob:quad} for all  $\lambda$ values combinations. For this example, we set $b=0.3393$. We began the optimization process for each such combination from the $\vx = \vy = [1 \quad 0 \quad 0]$ initial point, which is an approximate NE with $\varepsilon=b=0.3393$.
Here we can see that in most cases we can achieve a better approximation than the $0.3393$ of the initial point in the initial game.

We also tested our approach with a ``maximization version'' of the {\em Prisoner's Dilemma (PD)} game (i.e, in this version any player aims to maximizes her payoff, instead of minimizing her years in prison as in the most common PD variants). Our PD is defined as follows:
\begin{equation}
\label{game:additional_problem}
R = \begin{pmatrix}
2 & 0   \\
3 & 1  
\end{pmatrix},
\quad
C = \begin{pmatrix}
2 & 3  \\
0 & 1 
\end{pmatrix},
\end{equation}

\begin{figure}[h!]
    \centering
    \includegraphics[width=0.55\textwidth]{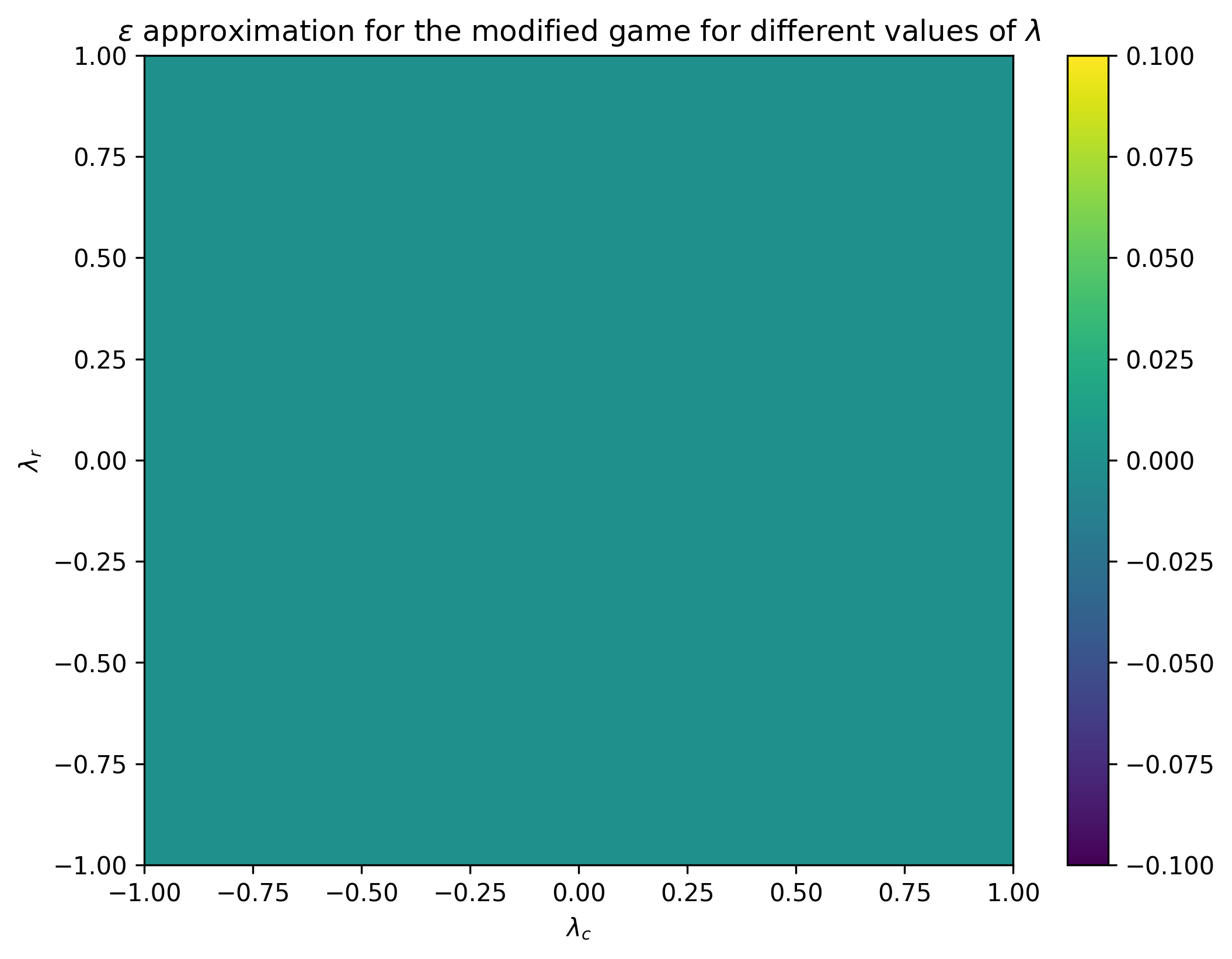}
    \caption{The $\epsilon$ approximation for our Prisoner's Dilemma (PD) game in its modified form. Averages over $20$ runs.}
\label{fig:additional_game_approximations}
\end{figure}

Fig.~\ref{fig:additional_game_approximations} shows the $\varepsilon$ approximation in the modified game of 
(\ref{game:additional_problem}) after converging to a strategy profile $(\vx,\vy)$ (following the same simulations process described for the Penalty Game above---starting again from a random initial strategy profile $(\vx, \vy)$).
Notice that excellent $\epsilon$-NE approximations (teal-colored region) are achieved for any combination of $\lambda$ values. 
This implies that in the PD's case an exact Nash equilibrium can be computed by the players for any combination of different behavioural policies ($\lambda$) of the players. This example gives insights for further research on the properties of games for which such a phenomenon holds.

Now, the following experiment tests Algorithm~\ref{prob:ts_problem} in the first motivating example (Example of Equation~\ref{game:ts_problem}). 
By running this alternating procedure, we simultaneously optimize the strategy and the ``behavioural'' profile of the two agents. 
We remind the reader that Algorithm~\ref{prob:ts_problem} alternatively modifies the $\lambda$ values and the strategies of the agents, progressively minimizing the regret  $\varepsilon$ (i.e., improving the approximation).
As such, we expect $\varepsilon$ to have a descending trajectory until it converges to a new value.

\begin{figure}
    \centering
    \includegraphics[width=0.6\linewidth]{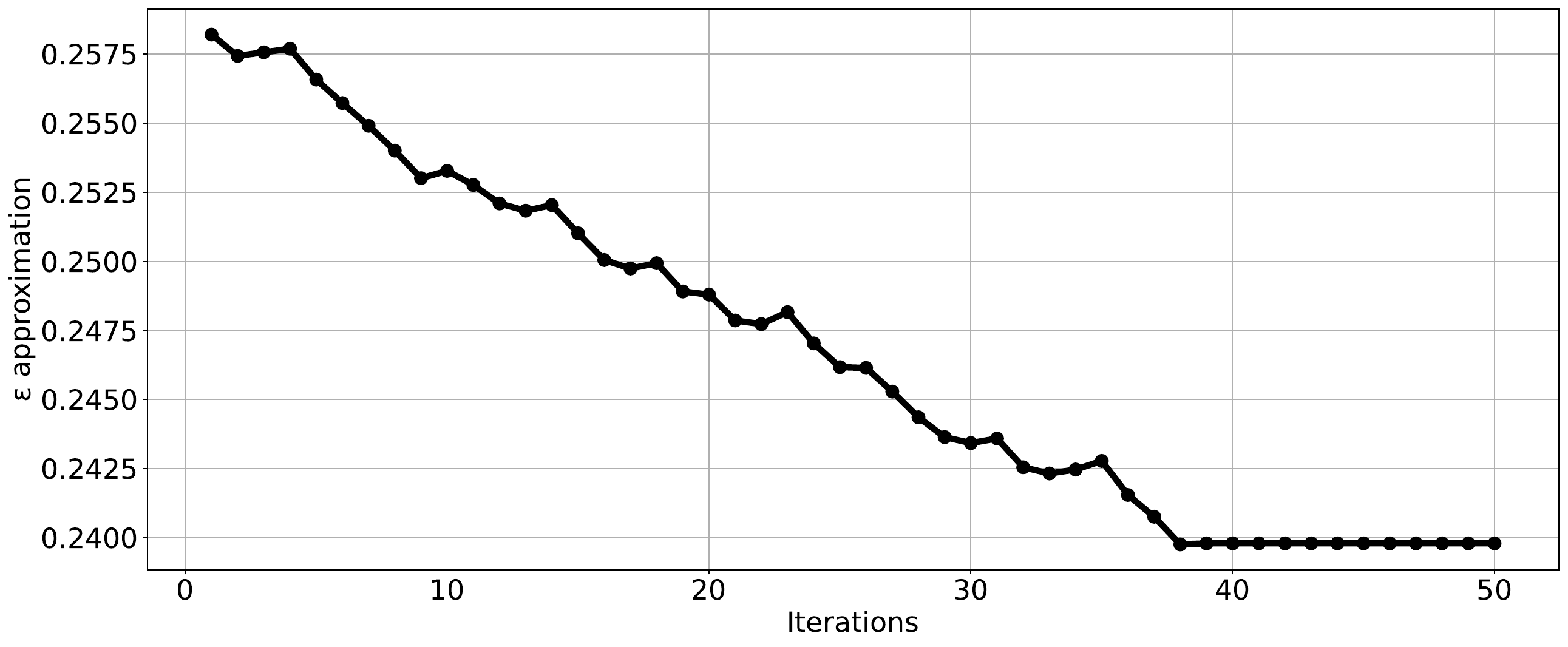}
    \caption{Alg.~\ref{prob:ts_problem} approximation for the example of Eq.~\ref{game:ts_problem}.}
\label{fig:motivation_example_1}
\end{figure}

Specifically, we ran Algorithm~\ref{prob:ts_problem} for the game presented in Equation~\ref {game:ts_problem}, for a number of $h=100$ iterations.
The algorithm initiates with $(\lambda_R, \lambda_C)=(0, 0)$ and strategy profile $\vx=\vy= [1 \quad 0 \quad 0]$. 
Figure~\ref{fig:motivation_example_1} shows the change of the $\varepsilon$ approximation for the modified game during a run of Algorithm~\ref{prob:ts_problem}. 
As observed, the approximation of the modified game indeed decreases as the algorithm fine-tunes the optimization variables appropriately. After a few iterations, it converges to a low approximation for the modified game.

Our experiments demonstrate some interesting phenomena. First, in the tight example of the TS algorithm the introduction of altruism/spite can lead players to points that have very good approximation, even $\varepsilon = 0$ (purple-blue in Fig.~\ref{fig:TS_approximations}). This shows that altruistic/spiteful behavior can potentially lead to more stable solutions, which is not always the case when 
players are selfish. Additionally, a better approximation can be achieved via Alg.~\ref{prob:ts_problem} (Fig.~\ref{fig:motivation_example_1}). 

\subsection{Opponent Modeling and Opponent Selection}
\label{sec:oppmodel}

We now turn our attention to settings 
with {\em uncertainty} regarding the $\lambda$ values of opponents. Specifically, consider a setting in which agents get the opportunity to observe the behaviour of potential opponents over some ``training'' period, during which it is matched against each one of them in a {\em repeated} mode. This allows agents to model their opponents and infer their $\lambda$'s. Intuitively, 
a learning agent is then able to ``select'' an opponent to face in a single-shot bimatrix game setting, 
given its own perception of what is beneficial to it (i.e., its own self-interest/altruism/spite point of view). 

Note that, contrary to what one could simplistically assume, it is not always the case that agents would prefer to play against an altruist. Consider, e.g., the $\varepsilon$ approximation in the TS tight example \cite{CDHLL23} (Fig.~\ref{fig:TS_approximations}). An agent with $\lambda = 0$ achieves a lower approximation when playing against a more spiteful ($\lambda < 0$) opponent. Thus, 
specific $\lambda$-opponent choice depends on the particular game setting.

We now detail our experimental setup. 
We have a total of $101$ agents in our setting.
For simplicity, we assume only one learning agent $i$, with a $\lambda_i = 0$,
known to all other agents $j$. 
The (100 in number) $\lambda_j$ values of the $j$ agents, are sampled from a uniform random distribution between $[-1,1]$. 
Then, $i$ plays $10^4$ times against each other agent $j$ in a (symmetric)  bimatrix game whose $R$ is randomly created from a 
Gaussian with $\mu = 0.5$ and $\sigma^2 = 0.2$, while $C=R^T$.
In this one-shot game, each agent has $15$ actions available. 

Each $j$ agent calculates at the start of the game its strategy profile $\vy^j$. This strategy is calculated by solving Problem~\ref{prob:quad2} 
given the known $\lambda$ values---$j$ knows both $\lambda_j$ and $\lambda_i$. Then, in each round, it chooses an action according to its $\vy^j$.
At the end of each round, agents observe the selected actions.

Now, agent $i$ has no knowledge of the $\lambda_j$ of any other agent $j$, and intends to learn it.
In principle, $i$  can use any opponent modeling
process of its choosing (e.g., a Bayesian beliefs updating one). 
Here it uses a simple process inspired by {\em fictitious play}~\cite{fudenberg98}: during the course of each repeated game, it maintains an {\em empirical distribution} that models the frequency of $j$'s actions.
Let us call this distribution $\hat{\vy}^{ij}$ ( $i$'s estimate of $j$'s strategy $\vy^j$). 

Then, agent $i$ (whose $\lambda_i$ is $0$) iteratively solves
program~\ref{prob:quad2} for all $\lambda$ values between $[-1,1]$ with `step' equal  to $0.1$ (ex. $[-1, -0.9 \ldots0.9,1]$). Hence, it is able to calculate a strategy 
$\vy({\lambda})$ for every potential $\lambda$ ``behavioural type'' facing $\lambda_i=0$. Thus, $i$ is able to compare $\hat{\vy}^{ij}$ with every 
$\vy({\lambda})$ 
and infer 
a $\hat{\lambda}_j$ value for $j$---i.e., sets 
$\hat{\lambda}_j = \argmin_{\lambda}||\vy({\lambda}) - \hat{\vy}^{ij}||_1$. Ideally, $\hat{\lambda}_j$ is close to the actual $\lambda_j$ characterizing agent $j$.

Having estimated a 
$\hat{\lambda}_j$
for every agent $j$, 
agent $i$
can now select opponents with a specific ``behavioural type'' to challenge them in a new one-shot (or repeated) game. 
Since the bimatrix game during this experiment is random, there is no intuitive explanation on which opponents (and of which ``behavioural type'') the agent $i$ should select. 
Since $i$ 
has $\lambda=0$, 
it selects as an opponent
a $j$ whose $\hat{\lambda}_j$ value is such that 
$i$'s reward is maximized 
given $\vy(\hat{\lambda}_j)$ and $\vx(\lambda_i)$, with ${\lambda_i=0}$. In this particular game, $i$ selects an opponent with
$\hat{\lambda}_j = -0.5$, that is, an opponent perceived as rather spiteful.

\begin{table}[hbtp]
\centering
\caption{Average rewards during training (before opponent selection) and evaluation (after opponent selection).}

\begin{footnotesize}
\begin{tabular}{l c}
Training Reward            & 0.5172 \\
\midrule
Evaluation Reward vs $\hat \lambda_j = -0.5$ & 0.5451 \\
Eval. Reward vs Spiteful opponents $(\lambda_j < -0.5)$ & 0.5422 \\
Eval. Reward vs Altruistic opponents $(\lambda_j >0.5)$ & 0.5302 \\
\bottomrule
\end{tabular}
\end{footnotesize}
\label{tab:results_learning_lambdas}
\end{table}

Table~\ref{tab:results_learning_lambdas} shows the reward of agent $i$ during this experiment.
In more detail, it shows the average reward achieved during the training phase, calculated across all $100$ opponents, over $10^4$ repeated game rounds per opponent. During training, agent $i$ used a random strategy, while every opponent $j$ was selecting actions according to its actual $\vy$ strategy, i.e. the strategy prescribed by its ``behavioural type'' when playing against 
a $\lambda=0$ agent, with $\vy$ estimated via Program~\ref{prob:quad2}. 

 Moreover, Table~\ref{tab:results_learning_lambdas} shows $i$'s 
 evaluation
 reward 
 $\vx^T R \vy$,
 after selecting its $j$ opponent according to the inferred $\hat \lambda_j$ =$-0.5$ behavioural type, and where $\vy$ is the strategy profile of the actual $\lambda_j$ opponent.
 We note that $\hat{\lambda}_j$ coincided with the actual $\lambda_j$ of that opponent---i.e., $i$
's learning process was successful. Indeed,  $i$ managed to learn the actual $\lambda$'s of all the opponents.
 
 For interest, Table~\ref{tab:results_learning_lambdas} also shows evaluation reward (i) 
 against all ``spiteful'' agents with $\lambda_j < -0.5$;
 and (ii) 
 against all ``altruists'' with $\lambda_j > 0.5$.
 We see that it is indeed beneficial to $i$ to play against $\tilde{\lambda}_j = -0.5$ (and to a lesser extent, against other, more spiteful agents), in this particular game.

\subsection{Transfer of knowledge and transfer learning}

Our agents' ability to model their opponents and learn their altrustic/spiteful behaviour, as outlined in the experiments above, provides the potential 
for the {\em transfer of knowledge}
between game environemnts.
That is, the potential arises for an agent that has learned an opponent's  $\lambda$ denoting its level of altruism (or spite) in one particular game setting, to {\em exploit} this knowledge in a {\em distinct} game setting, potentially increasing its reward in that second game compared to what it would have been had it been ignorant of the actual opponent's $\lambda$ (and thus had effectively assumed it to be equal to zero as the SIH requires). For example, consider the game
\begin{equation*}
\label{game:example}
G = \begin{pmatrix}
4,2 \quad & 0,3\\
3,0 \quad & 1,1\\
\end{pmatrix}.
\end{equation*}
We can see that the only NE is the strategy profile with payoffs $(1,1)$. Let us now assume that the Row player has $\lambda_R = 0$ and {\em has learned} (possibly via training in a completely different game in the past) the $\lambda_C = 1$ of the Column player. Then, 
Row realizes that the actual game that will be played against Column, is

\begin{equation*}
\label{game:example}
G' = \begin{pmatrix}
4,6 \quad & 0,3\\
3,3 \quad & 1,2\\
\end{pmatrix}.
\end{equation*}

We can see that the unique NE in game $G'$ is the strategy profile with payoffs $(4,6)$. Thus, since Row  knows the $\lambda_C$ determining the respective strategy of the Column player, she will adapt her own strategy and  achieve a better payoff (i.e., a $4$ instead of a $1$).

Moreover, we point out that the opponent modeling process of Section~\ref{sec:oppmodel} can be used for {\em transfer learning} purposes, in the spirit of the namesake machine learning concept~\cite{pan2010}.
That is, in our problem domain, an agent can leverage its experience from learning, at least ``partially'' or with some degree of confidence, the $\lambda$ of an opponent in a game (e.g., via the process presented in Section~\ref{sec:oppmodel}), in order to accelerate its learning of that $\lambda$ value in a second, distinct to the first one, game setting (via the same learning process, or, in principle, via any learning process of its choosing).
For instance, transfer learning could be particularly useful in situations when the first game is a small one, while the second game is a significantly larger one (in terms of the action set and thus potentially the equilibrium strategy support set); i.e., one that would require a large number of iterations (rounds) in order to learn it.

 To show this, we conducted a small experiment using two random symmetric games: a $3 \times 3$ game, denoted as $A$, and a $10 \times 10$ game, denoted as $B$. The payoff matrices for both games were sampled from a normal distribution with $\mu = 0.5$ and $\sigma = 0.2$. 
Agent $i$ interacted with agent $j$ during game $A$ for a number of $X=3$ times, tracking the empirical distribution $\vy_A^{ij}$ ($i$'s 
estimate of $j$'s $\vy^j$ strategy in $A$). 
Then, $i$ inferred the type $\hat{\lambda}_j$ of agent $j$ using $\vy_A^{ij}$, in the same way as in Section~\ref{sec:oppmodel}. 
Consequently, using $\hat{\lambda}_j$, agent $i$ calculated 
a ``prior'' 
strategy profile 
for $j$
in game $B$ via solving Problem~\ref{prob:quad2}. 
 Agent $i$ 
 then 
 updated
 that
 strategy profile using $Y=53$ interactions in game $B$ until convergence---as such, 
 finally 
 inferring $j$'s type.
Now, another agent $k$ inferred $j$'s type via interacting $Z$ times {\em in game $B$ only}. The results were as follows: Both agent $i$ and $k$ inferred the correct $\lambda$ value for agent $j$. However, agent $k$ did so via interacting with $j$ for $Z = 706$ times in total, in game $B$ only;  while $i$ was able to do so via interacting with $j$ for a total of $X + Y = 56$ times only, during both games. 
This verifies the potential of transfer learning in our framework: 
 Agent $i$
required {\em an order of magnitude fewer interactions in game $B$ compared to $k$}, since, unlike agent $k$, it had obtained an informed prior about $j$
via interacting with it in $A$.

 \section{Related Work}

As mentioned in the introduction, there exist several models in economics that define various functional forms of preferences and utility, aiming to depart from the self-interested hypothesis (SIH)~\cite{levine1998,bolton00,sobel2005}. As a result several manifestations of spite and/or altruism emerge in these models.
In our work, we chose to build upon the rather intuitive definition of altruism appearing in~\cite{MM08}. That work provides a refinement of the NE which incorporates a small degree of reciprocal altruism, so as to encourage ``friendly'' behaviour and fend-off instability against small perturbations in the NE payoffs---similarly to a trembling-hand equilibrium~\cite{Selten1975}, but with an altruistic twist. In our work, we are inspired by their definition of altruism to define spite in bimatrix games; and then tackle these notions from an algorithmic game theoretic point of view, while also taking steps towards a (machine) learning treatment of this topic.

There are many game-theoretic works that consider altruism either in cooperative~\cite{kerkmann2024altruism} or in non-cooperative settings~\cite{Rothe_2021,ashlagi2008,anagnostopoulos2013}. The latter, however, either study specific game properties or player behaviour, such as specific forms of equilibria or pure NE in specific classes of games, e.g., congestion games or graph-restricted settings; but do not focus on bimatrix games nor study approximate NE as we do in our work. In most cases, existing work focuses on price of anarchy, price of stability, and (in-)efficiency of equilibria or mechanism design-relating questions~\cite{Rothe_2021,brokkelkamp2022greater,bilo2013,ChenKKS14}. Apt and Sch{ä}fer~\cite{apt-shafer-2014}, for instance, define and study the ``selfishness level of strategic games''. In essence, they
 measure the smallest adjustment needed to be performed to player payoffs so that the socially optimal outcome is also a stable Nash equilibrium. Their work does not question the SIH per se, but effectively determines how much players would need to value the collective good so that a socially desirable outcome can be achieved via their self-interested choices. We have not so far studied such questions, but in principle they could be incorporated in our model---for instance, it could be interesting to study the form and degree of incentives that should be provided to individuals in order for them to alter their degree of altruism/spite.

\section{Conclusions and future work}
In our work in this paper, we move away from the SIH to provide both a  theoretical and an experimental treatment of altruism and spite in the fundamental class of bimatrix games.
We provide 
theoretical results regarding (approximate) NE under altruism/spite, showing that some better that the state-of-the-art approximations can be computed in poly-time when altruism or spite are at play;
 and contribute a PGD-based algorithm that optimizes towards a regret-minimizing strategies/behaviours profile. Furthermore, we 
explored the potential of opponent modeling and learning given uncertainty over the levels of altruism.

In future work, we intend to tackle  {\em $n$-player normal form}, or  {\em polymatrix games}~\cite{DFSS17}.
Moreover, it would be interesting to add stochasticity in the game outcomes,
requiring
the learning of the reward matrices alongside the $\lambda$s, in a reinforcement learning fashion. 
Furthermore, we intend to explore the effects that 
{\em non-stationary} $\lambda$ values have to convergence to (approximate)
NE 
More generally, it would 
be interesting to study the convergence of evolutionary learning/replicator dynamics in this context---e.g., identifying the behaviour dynamics' steady states that emerge after repeated play in large, learning populations~\cite{LNAI18-Tuyls,fudenberg98}.
As such, our work can be viewed as a stepping stone towards integrating AGT 
solutions
with ML research---including work towards revealing the implications of bounded rationality in agent decision making.
\begin{credits}
\subsubsection{\ackname} The research described in this paper was carried out within the framework of
the National Recovery and Resilience Plan Greece 2.0, funded by the European Union - NextGenerationEU (Implementation Body: HFRI. Project name:
DEEP-REBAYES. HFRI Project Number 15430).
\end{credits}

\bibliographystyle{splncs04}
\bibliography{sample}

\end{document}